\documentclass[runningheads]{llncs}
\usepackage{calc}
\usepackage{algpseudocode}
\usepackage{algorithm}
\usepackage{algpseudocode}
\usepackage{blindtext}
\let\llncssubparagraph\subparagraph
\let\subparagraph\paragraph
\usepackage[]{titlesec}
\let\subparagraph\llncssubparagraph

\titleformat{\paragraph}
{\normalfont\normalsize\bfseries}{\theparagraph}{1em}{}
\titlespacing*{\paragraph}
{0pt}{3.25ex plus 1ex minus .2ex}{1.5ex plus .2ex}
\usepackage{tikz}
\usetikzlibrary{decorations.pathreplacing}
\usetikzlibrary{decorations.markings}
\tikzstyle{vertex}=[circle, draw, inner sep=0pt, minimum size=3pt]

\usepackage{graphicx}
\usepackage{amsmath}
\usepackage{amsfonts}
\usepackage{xcolor}
\usepackage{amssymb}
\usepackage{graphicx}
\usepackage{tikz}
\usetikzlibrary{circuits.logic.US}
\usetikzlibrary{positioning}
\usetikzlibrary{matrix}
\usepackage{graphicx}
\usepackage{amsmath}
\newcommand{\boundellipse}[3]
{(#1) ellipse (#2 and #3)
}
\definecolor{magenta}{rgb}{0.8, 0.0, 0.8}
\definecolor{cyan}{rgb}{0.0, 1.0, 1.0}
\definecolor{green1}{rgb}{0.1, 0.6, 0.01}
\definecolor{green}{rgb}{0.11, 0.35, 0.02}
\definecolor{brown}{rgb}{0.65, 0.16, 0.16}
\definecolor{cadetgrey}{rgb}{0.57, 0.64, 0.69}

\usepackage{xspace}
\newcommand{\cO}{\mathcal{O}\xspace}
\newcommand{\csO}{\mathcal{O}^{\star}\xspace}
\newcommand{\vc}{\mathsf{vc}}

\newcommand{\mmfvs}{{\sc Max Min FVS}\xspace}

\newtheorem{red}{\bf Reduction EMMFVS}

\title{Maximum Minimal Feedback Vertex Set: A Parameterized Perspective}
\author{Ajinkya Gaikwad\inst{1}
\and Hitendra Kumar \inst{1}
\and Soumen Maity\inst{1}
\and  \\Saket Saurabh\inst{2,3} 
\and Shuvam Kant Tripathi\inst{1} 
}
\authorrunning{A.\,Gaikwad et al.}
\institute{Indian Institute of Science Education and Research, Pune, India 
\and
The Institute of Mathematical Sciences, Chennai, India \and University of Bergen, Bergen, Norway\\
\email{\texttt{ajinkya.gaikwad@students.iiserpune.ac.in}};
\email{\texttt{soumen@iiserpune.ac.in}};
\email{\texttt{saket@imsc.res.in}}\\
}

\begin{document}

\maketitle          
\begin{abstract}In this paper we study a maximization version of 
the classical {\sc Feedback Vertex Set (FVS)} problem, namely, the {\sc Max Min FVS} problem, in the realm of parameterized complexity. In this problem, given an undirected graph $G$, a positive integer $k$, the question is to check whether $G$ has a minimal feedback vertex set of size at least $k$.  We obtain following results for {\sc Max Min FVS}. 
\begin{enumerate}
    \item We first design  a fixed parameter tractable (FPT)  algorithm for  {\sc Max Min FVS}  running in time $10^kn^{\mathcal{O}(1)}$.
    \item Next, we  consider the problem parameterized by the vertex cover number of the input graph (denoted by 
    $\vc(G)$), 
    and design an algorithm with running time $2^{\cO(\vc(G)\log \vc(G))}n^{\mathcal{O}(1)}$. We complement this result by showing that the problem parameterized by $\vc(G)$ does not admit a polynomial 
compression unless coNP $\subseteq$ NP/poly.
\item Finally, we give an FPT-approximation scheme 
(fpt-AS)  parameterized by $\vc(G)$. That is, we design an algorithm that for every $\epsilon >0$, runs in time $2^{\cO\left(\frac{vc(G)}{\epsilon}\right)} n^{\mathcal{O}(1)}$ and returns a minimal feedback vertex set of size at least $(1-\epsilon){\sf opt}$. 
\end{enumerate}

 \keywords{Parameterized Complexity \and FPT \and vertex cover}
\end{abstract}

\section{Introduction}
{\sc Feedback Vertex Set (FVS)} together with {\sc Vertex Cover} are arguably the two most well studied problems in parameterized complexity. In FVS, we are given  an undirected graph $G$, a positive integer $k$, and  the question is to check whether $G$ has a vertex set $S$ of size at most $k$ such that $G-S$ is a forest (an acyclic graph).  The set $S$ is called {\em feedback vertex set} or {\em fvs} in short.  
Downey and Fellows \cite{Downey_First_algo}, and Bodlaender \cite{Bodlaender} proposed the first two fixed parameter tractable (FPT) algorithms with the running time $\csO(2^{\cO(k \log k)})$\footnote{The notation $\csO$ hides the polynomial factor in the running time.}. 
After a long series of improvements, the current champion algorithms are as follows. The fastest known randomized algorithm is given by Li and Nederlof~\cite{LiN20} and runs in time $\csO(2.7^k)$, and the fastest known deterministic algorithm, given by  Iwata and  Kobayashi~\cite{IwataY19}, runs in time $\csO(3.460^k)$.  Several minimization variants of FVS have been studied in the literature such as finding a set $S$ such that $G-S$ is acyclic and $G[S]$ is connected or  $G[S]$ is an independent set. In this paper we consider a (not so well studied)  maximization version of FVS, namely \mmfvs. 
A set $S$ is called a {\em minimal fvs}, if $S$ is an fvs and for every $v\in S$, $S\setminus \{v\}$ is not an fvs. That is, no proper subset of $S$ is an fvs. It is not hard to see that if $S$ is a minimal FVS, then every $u\in S$ has a 
\emph{private cycle}, that is, there exists a cycle in $G[(V(G)\setminus S)\cup \{u\}]$, which goes through $u$. Now we are ready to define the problem  formally. 
    \begin{center}
    \fbox
    {\begin{minipage}{33.7em}\label{FFVS }
       {\sc Max Min FVS}\\
        \noindent{\bf Input:} An undirected graph $G=(V,E)$ and an  integer $k \in \mathbb{N}$.
    
        \noindent{\bf Question:}  Does $G$ admit a minimal FVS of size greater than 
        or equal to $k$? 
    \end{minipage} }
\end{center}
The graph parameter we discuss in this paper is  vertex cover number. 
\begin{definition}\rm 
A set $C\subseteq V$ is a \emph{vertex cover} of $G=(V,E)$ if each edge $e\in E$ has at least one endpoint 
 in $C$. The minimum size of a vertex cover in $G$ is the {\it vertex cover number} of $G$, denoted by $\vc(G)$.
 \end{definition}   

Lately, finding a large size minimal solution has attracted a lot of attention from the perspective of the Approximation Algorithms and Parameterized Complexity. Boria et al.~\cite{BoriaCP15} proved that for any constant $\epsilon >0$, the optimization version of  {\sc Max Min Vertex Cover} is inapproximable within ratios $\cO(n^{0.5-\epsilon})$, unless P=NP. They complement this result by proving that {\sc Max Min Vertex Cover} is approximable within ratio $\cO(n^{0.5})$ in polynomial time. This is in sharp contrast to the approximability of the classical {\sc Vertex Cover} problem, for which an easy factor $2$-approximation exists.  This becomes even more interesting when we consider the optimization version of \mmfvs.  The {\sc Max Min FVS} problem
was first considered by Mishra and Sikdar \cite{Sounaka}, who showed that the problem does not admit an
$n^{0.5-\epsilon}$ approximation (unless P=NP), and that it remains APX-hard
even when the input graph is of degree at most 9.  Dublois et al.~\cite{DubloisHGLM22} improved upon this by showing the first non-trivial polynomial time approximation for \mmfvs with a ratio of $\cO(n^{\frac{2}{3}})$, as well as a matching hardness of approximation bound of $n^{\frac{2}{3}-\epsilon}$. Apart from these two problems, there are many other classical optimization problems that have recently been studied in the {\sc MaxMin} or {\sc MinMax} framework, such as {\sc Max Min Separator} \cite{HANAKA2019294} and {\sc Max Min Cut} \cite{Eto2019ParameterizedAF}.


 In the realm of parameterized complexity,  Zehavi~\cite{Zehavi17} studied {\sc Max Min Vertex Cover} -- 
 find a {\em minimal vertex cover} of size at least $k$, if exists -- and designed an algorithm with running time $\csO(2^{\vc(G)})$, which is "almost optimal" unless Strong Exponential Time Hypothesis fails. For \mmfvs, Dublois et al.~\cite{DubloisHGLM22} obtained a polynomial kernel of size $\cO(k^3)$. That is, they design a polynomial time algorithm that given an instance $(G,k)$ returns an equivalent instance $(G',k')$ such that $k'\leq k$ and $|V(G')+E(G')|\leq \cO(k^3)$. This result is the starting point of our work. 
 There are results about kernelization of {\sc Max Min Vertex Cover} and {\sc Max Min FVS} in 
 \cite{IPEC2021new}.
In particular, they proved that {\sc Max Min VC} parameterized by vertex cover number does not admit a polynomial kernel. This result is  related to the kernelization of {\sc Max Min FVS }
 parameterized by vertex cover of our work.



\subsection{Preliminaries} We only consider finite 
undirected graphs without loops or multiple edges, and we denote 
an edge between two vertices $u$ and $v$ by $(u,v)$. 
A subgraph $H$ of a graph $G$ is \emph{induced} if $H$ can be obtained from $G$ by deleting a set of vertices $D=V(G)\setminus S$, and we denote $H=G[S]$. 
For a graph $G$ and a set $S\subseteq V(G)$, we use the notation $G-S = G[V (G) \setminus S]$, and  for a vertex $v \in V (G)$, we abbreviate 
$G \setminus \{v\}$ as $G-v$. The {\it (open) neighbourhood} $N_G(v)$ of a vertex 
$v\in V(G)$ is the set $\{u~|~(u,v)\in E(G)\}$. The {\it closed neighbourhood} $N_G[v]$ of a vertex $v\in V(G)$ is the set
$\{v\} \cup N_G(v)$.  The {\it degree} of $v\in V(G)$ is $|N_G(v)|$ and denoted by $d_G(v)$. We use $d_S(v)$ to denote the degree of vertex $v$ in $G[S]$. 
For an integer $n \geq 1$, we let $[n]$ be the set containing all 
integers $i$ with  $1 \leq  i \leq  n$. In a graph $G$, \emph{contraction} of an edge $e=(u,v)$  is the replacement of $u$ and $v$ with a single vertex such that edges incident to the new vertex are the edges other than $e$ that were incident with $u$ or $v$. The resulting graph, denoted $G/e$, has one less edge than $G$.  We refer to Appendix A and \cite{marekcygan,Downey} for  details on parameterized complexity.

\subsection{Our results and methods}  Using, the polynomial kernel, of size $\cO(k^3)$, of  Dublois et al.~\cite{DubloisHGLM22}, we can design an FPT algorithm for \mmfvs running in time $\csO(2^{\cO(k^3)})$ as follows. For every vertex subset $S$ of size at least $k'$ of $V(G')$ test whether $S$ is a minimal fvs or not. If we succeed for any $S$, we have that $(G,k)$ is a yes instance, else, it is a no instance. As our first result we improve upon this result and obtain the following result. 

\begin{theorem}
\label{thmintro:mmfvs}
\mmfvs can be solved in time $\csO(10^k)$. 
\end{theorem}

\begin{figure}[ht]
    \centering
    \begin{tikzpicture}[scale=0.5]
    \node[circle,draw,fill=red, inner sep=0 pt, minimum size=0.15cm]	(p1) at (-14,14) [label=left:$x$]{};
    \node[circle,draw,fill=black, inner sep=0 pt, minimum size=0.15cm]	(p2) at (-14,11.5) [label=left:$y$]{};
    \node[circle,draw,fill=pink, inner sep=0 pt, minimum size=0.15cm]	(p3) at (-12,15) [label=right:$a$]{};
    \node[circle,draw,fill=pink, inner sep=0 pt, minimum size=0.15cm]	(p4) at (-12,13.5) [label=right:$b$]{};
    \node[circle,draw,fill=pink, inner sep=0 pt, minimum size=0.15cm]	(p5) at (-12,12) [label=right:$c$]{};
    \node[circle,draw,fill=pink, inner sep=0 pt, minimum size=0.15cm]	(p6) at (-12,10.5) [label=right:$d$]{};
    
    \draw(p1)--(p3);
    \draw(p1)--(p4);
    \draw(p1)--(p5);
    \draw(p1)--(p6);
    \draw(p2)--(p3);
    \draw(p2)--(p4);
    \draw(p2)--(p5);
    \draw(p2)--(p6);
    \draw(p1)--(p2);
    
    \end{tikzpicture}
    \caption{Graph $G$ with ${\sf fvs}(G)=1$ and ${\sf opt}(G)=4$}
    \label{FVS vs MFVS}
\end{figure}
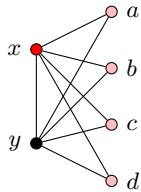

Proof of Theorem~\ref{thmintro:mmfvs} follows the strategy used for designing an iterative compression based FPT algorithm for FVS.  Let ${\sf fvs}(G)$ denote the minimum size of a feedback vertex set of $G$, and let
${\sf opt}(G)$ denote the maximum size of a minimal feedback vertex set of $G$. 
Clearly, ${\sf fvs}(G) \leq {\sf opt}(G)$. The gap between these quantities can be arbitrary large -- as 
shown in Figure \ref{FVS vs MFVS}. Here,  ${\sf fvs}(G)=1$, while ${\sf opt}(G)=|V(G)|-2$. Further, observe that the same example also shows that the gap  between  $\vc(G)$ and ${\sf opt}(G)$ can be arbitrary large. Here,  $\vc(G)=2$, while ${\sf opt}(G)=|V(G)|-2$.  The discussion above implies that for \mmfvs both  ${\sf fvs}(G)$ and $\vc(G)$ are interesting parameters to consider. For our second result, we consider \mmfvs parameterized by $\vc(G)$ and obtain the following result.

\begin{theorem}
\label{thmintro:mmfvsvc}
\mmfvs can be solved in time $\csO(2^{\cO(\vc(G) \log \vc(G))})$. 
\end{theorem}

The starting point of the algorithm is based on the natural partitioning ideas. However, to complete the algorithm we need to design an algorithm for {\sc Induced Forest Isomorphism} parameterized by $\vc(G)$, which could be of independent interest. This algorithm is the bottleneck in designing an algorithm for  \mmfvs running in time  $\csO(2^{\cO(\vc(G)\log \vc(G))})$. We complement this result by showing that \mmfvs parameterized by  $\vc(G)$ does not admit a polynomial 
compression unless coNP $\subseteq$ NP/poly. 
Note that $\vc(G)$ can be much larger than ${\sf opt}(G)$, for example in a cycle, and so 
Theorem \ref{thmintro:mmfvs} is not implied by Theorem \ref{thmintro:mmfvsvc}.  Finally, we show that if we allow a ``small loss'' then we can improve upon Theorem~\ref{thmintro:mmfvsvc}. That is, we design an FPT-approximation algorithm for \mmfvs parameterized by  
$\vc(G)$. 

\begin{theorem}
\label{thmintro:mmfvsvcapx}
Let $\epsilon >0$ be a fixed constant. Then, there exists an algorithm for 
\mmfvs, that runs in time
 $2^{\cO(\frac{\vc(G)}{\epsilon})}n^{\cO(1)}$ and returns a minimal feedback vertex set of size at least $(1 - \epsilon){\sf opt}(G)$. \\
\end{theorem}

\section{FPT algorithm parameterized by solution size}\label{FPT-Algo}
In this section we design an FPT algorithm for {\sc Max Min FVS}. First we give an algorithm for {\sc Extension Max Min FVS}. 
We start by defining {\sc Extension Max Min FVS}. In this problem, as input instance
$I=(G,W_{1},W_{2},k)$, we are
given an undirected graph $G$, an integer $k$ and a minimal feedback vertex set  $W=W_1\cup W_2$ in $G$.  We are also given a partition 
$(W_1,W_2)$ of the vertices of $W$. 
The objective is to decide if $G$ has a 
minimal feedback vertex set $S$ such that $W_{1} \subseteq S$, $W_2\cap S = \emptyset$ and 
$ |S\setminus W_{1}| \geq k$, or correctly
conclude that no such minimal feedback vertex set exists. We give an algorithm for 
{\sc Extension Max Min FVS} running in time ${3}^{k+\gamma(I)}n^{\cO(1)}$, where $\gamma(I)$
is the number of connected components of $G[W_2]$.

\subsection{An algorithm for {\sc Extension Max Min FVS} }
Let $I=(G,W_{1},W_{2},k)$ be an instance of {\sc Extension Max Min FVS} and let 
$H=G-W$ where $W=W_1\cup W_2$ and $W_1\cap W_2=\emptyset$. We first give some reduction
rules to simplify the input instance. 

\begin{red}\label{Red1}\rm 
If $G-W_{1}$ has a vertex $v$ of degree at most 1, remove it from the graph. 
\end{red}
    Reduction EMM-FVS \ref{Red1} is safe, because for a given instance $(G,W_{1},W_{2},k)$ of {\sc Extension Max Min FVS}, if the graph $G-W_{1}$ has a vertex of degree at most one, then this vertex is not part of any cycle in $G-W_{1}$ . Thus, its removal does not change the solution.

\begin{red}\label{Red2}\rm
If there is a vertex $v$ in $H$ such that $G[W_{2} \cup \{v\}]$ contains a cycle, then include $v$ in $W_{1}$, and decrease the parameter by 1.  
That is, the new instance is $(G,W_{1}\cup \{v\},W_{2},k-1)$.
\end{red}
Reduction MMFVS \ref{Red2} is safe. Suppose $G[W_{2} \cup \{v\}]$ contains a cycle $C$.  As the solution here has to be disjoint from $W_{2}$, the only way to destroy $C$ is to include $v$ in the solution.

\begin{red}\rm 
 If $(u,v) \in E(G)$ such that $N(u) \cap N(v)= \emptyset$, $d(v)=d(u)=2$ and $u,v\notin W$, then contract $(u,v)$.  That is, the new instance is $(G/(u,v),W_{1},W_{2},k)$.
\end{red}
Reduction Rule EMMFVS 3 is safe as any minimal feedback vertex set contains at most one of 
$u$ and $v$.  This reduction rule is inspired from \cite{DubloisHGLM22} and a formal proof 
of this rule is given in \cite{DubloisHGLM22}.
Furthermore, all reductions can be applied in polynomial time. 

\begin{lemma}
{\sc Extension Max Min FVS} can be solved in time $3^{k+\gamma(I)} n^{\cO(1)}$.
\end{lemma}

\proof Let $(G,W_{1},W_{2},k)$ be the input instance.  If $G[W_{2}]$ is not a forest then we return that $(G,W_{1},W_{2},k)$ is a no-instance. So from now onward we assume that $G[W_{2}]$ is a forest. 
We follow a branching technique with a measure function $\mu$.
For instance $I=(G,W_{1},W_{2},k)$, we define its measure $$\mu(I)= k+\gamma(I)$$
where $\gamma(I)$ is the number of connected components of $G[W_{2}]$.
The algorithm first applies Reduction EMMFVS 1, EMMFVS 2, and EMMFVS 3 exhaustively. 
For clarity we denote the reduced instance by $(G,W_{1},W_{2},k)$.
Since $W$ is a feedback vertex set, $H=G-W$ is a forest.  Thus $H$ has a vertex of degree at most 1.
In each tree of the forest $H$, arbitrarily pick one of its vertices as the root.  Now we  focus on a deepest leaf $v$ of any tree in $H$. Clearly $v$ has at least one neighbour in $W_{2}$, otherwise Reduction EMMFVS 1 would have been applied. We distinguish  two cases based on the number of neighbours of $v$ in $W_{2}$.\\
 
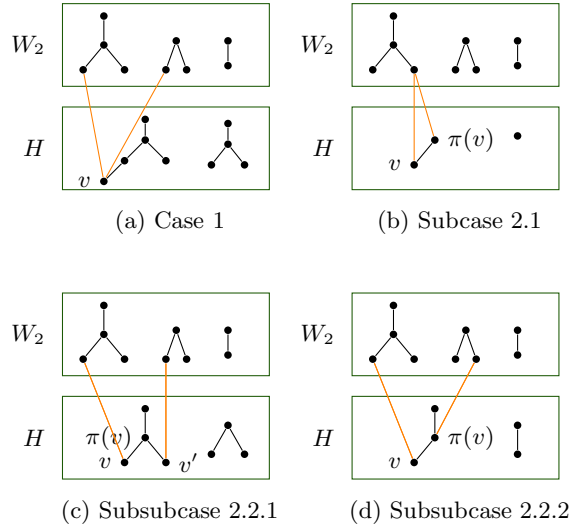
\begin{figure}
    \centering
   \begin{tikzpicture}[scale=0.55]
    
\draw[green] (-1,0) rectangle (4,2);  
\draw[green] (6,0) rectangle (11,2);      
\draw[green] (-1,2.5) rectangle (4,4.5);  
\draw[green] (6,2.5) rectangle (11,4.5);

\draw[green] (-1,7) rectangle (4,9);  
\draw[green] (6,7) rectangle (11,9);      
\draw[green] (-1,9.5) rectangle (4,11.5);  
\draw[green] (6,9.5) rectangle (11,11.5);   
    
\node[circle, fill=black, inner sep=0 pt, minimum size=0.1cm] (x1) at (1,1.7) []{};  
\node[circle, fill=black, inner sep=0 pt, minimum size=0.1cm] (x2) at (1,1) [label=left:$\pi(v)$]{};  
\node[circle, fill=black, inner sep=0 pt, minimum size=0.1cm] (x3) at (0.5,0.4) [label=left:$v$]{};
\node[circle, fill=black, inner sep=0 pt, minimum size=0.1cm] (x4) at (1.5,0.4) [label=right:$v'$]{}; 
\node[circle, fill=black, inner sep=0 pt, minimum size=0.1cm] (x5) at (3,1.3) []{};  
\node[circle, fill=black, inner sep=0 pt, minimum size=0.1cm] (x6) at (3.4,0.6) []{}; 
\node[circle, fill=black, inner sep=0 pt, minimum size=0.1cm] (x7) at (2.6,0.6) []{}; 
\node[circle, fill=black, inner sep=0 pt, minimum size=0.1cm] (x8) at (0,4.2) []{};  
\node[circle, fill=black, inner sep=0 pt, minimum size=0.1cm] (x9) at (0,3.5) []{};  
\node[circle, fill=black, inner sep=0 pt, minimum size=0.1cm] (x10) at (-.5,2.9) []{};
\node[circle, fill=black, inner sep=0 pt, minimum size=0.1cm] (x11) at (.5,2.9) []{}; 
\node[circle, fill=black, inner sep=0 pt, minimum size=0.1cm] (x12) at (1.5,2.9) []{};
\node[circle, fill=black, inner sep=0 pt, minimum size=0.1cm] (x13) at (2,2.9) []{};
\node[circle, fill=black, inner sep=0 pt, minimum size=0.1cm] (x14) at (1.75,3.6) []{};

\node[circle, fill=black, inner sep=0 pt, minimum size=0.1cm] (x15) at (3,3.6) []{};
\node[circle, fill=black, inner sep=0 pt, minimum size=0.1cm] (x16) at (3,3) []{};

\draw(x1)--(x2);
\draw(x3)--(x2);
\draw(x2)--(x4);
\draw(x5)--(x6);
\draw(x5)--(x7);
\draw(x8)--(x9);
\draw(x9)--(x10);
\draw(x9)--(x11);
\draw(x12)--(x14);
\draw(x14)--(x13);
\draw(x15)--(x16);

\node[circle, fill=black, inner sep=0 pt, minimum size=0.1cm] (y1) at (8,1.7) []{};  
\node[circle, fill=black, inner sep=0 pt, minimum size=0.1cm] (y2) at (8,1) [label=right:$\pi(v)$]{};  
\node[circle, fill=black, inner sep=0 pt, minimum size=0.1cm] (y3) at (7.5,0.4) [label=left:$v$]{};
\node[circle, fill=black, inner sep=0 pt, minimum size=0.1cm] (y5) at (10,1.3) []{};  
\node[circle, fill=black, inner sep=0 pt, minimum size=0.1cm] (y7) at (10,0.6) []{}; 
\node[circle, fill=black, inner sep=0 pt, minimum size=0.1cm] (y8) at (7,4.2) []{};  
\node[circle, fill=black, inner sep=0 pt, minimum size=0.1cm] (y9) at (7,3.5) []{};  
\node[circle, fill=black, inner sep=0 pt, minimum size=0.1cm] (y10) at (6.5,2.9) []{};
\node[circle, fill=black, inner sep=0 pt, minimum size=0.1cm] (y11) at (7.5,2.9) []{}; 
\node[circle, fill=black, inner sep=0 pt, minimum size=0.1cm] (y12) at (8.5,2.9) []{};
\node[circle, fill=black, inner sep=0 pt, minimum size=0.1cm] (y13) at (9,2.9) []{};
\node[circle, fill=black, inner sep=0 pt, minimum size=0.1cm] (y14) at (8.75,3.6) []{};
\node[circle, fill=black, inner sep=0 pt, minimum size=0.1cm] (y15) at (10,3.6) []{};
\node[circle, fill=black, inner sep=0 pt, minimum size=0.1cm] (y16) at (10,3) []{};

\draw(y1)--(y2);
\draw(y3)--(y2);
\draw(y5)--(y7);
\draw(y8)--(y9);
\draw(y9)--(y10);
\draw(y9)--(y11);
\draw(y12)--(y14);
\draw(y14)--(y13);
\draw(y15)--(y16);

\path 
(y3) [orange] edge (y10)
(y2) [orange] edge (y13)
(x3) [orange] edge (x10)
(x4) [orange] edge (x12);

\node[circle, fill=black, inner sep=0 pt, minimum size=0.1cm] (z1) at (1,8.7) []{};  
\node[circle, fill=black, inner sep=0 pt, minimum size=0.1cm] (z2) at (1,8.2) []{};  
\node[circle, fill=black, inner sep=0 pt, minimum size=0.1cm] (z3) at (0.5,7.7) []{};
\node[circle, fill=black, inner sep=0 pt, minimum size=0.1cm] (z17) at (0,7.2) [label=left:$v$]{};
\node[circle, fill=black, inner sep=0 pt, minimum size=0.1cm] (z4) at (1.5,7.7) []{}; 
\node[circle, fill=black, inner sep=0 pt, minimum size=0.1cm] (z5) at (3,8.1) []{};  
\node[circle, fill=black, inner sep=0 pt, minimum size=0.1cm] (z18) at (3,8.6) []{};  
\node[circle, fill=black, inner sep=0 pt, minimum size=0.1cm] (z6) at (3.4,7.6) []{}; 
\node[circle, fill=black, inner sep=0 pt, minimum size=0.1cm] (z7) at (2.6,7.6) []{}; 
\node[circle, fill=black, inner sep=0 pt, minimum size=0.1cm] (z8) at (0,11.2) []{};  
\node[circle, fill=black, inner sep=0 pt, minimum size=0.1cm] (z9) at (0,10.5) []{};  
\node[circle, fill=black, inner sep=0 pt, minimum size=0.1cm] (z10) at (-.5,9.9) []{};
\node[circle, fill=black, inner sep=0 pt, minimum size=0.1cm] (z11) at (.5,9.9) []{}; 
\node[circle, fill=black, inner sep=0 pt, minimum size=0.1cm] (z12) at (1.5,9.9) []{};
\node[circle, fill=black, inner sep=0 pt, minimum size=0.1cm] (z13) at (2,9.9) []{};
\node[circle, fill=black, inner sep=0 pt, minimum size=0.1cm] (z14) at (1.75,10.6) []{};
\node[circle, fill=black, inner sep=0 pt, minimum size=0.1cm] (z15) at (3,10.6) []{};
\node[circle, fill=black, inner sep=0 pt, minimum size=0.1cm] (z16) at (3,10) []{};

\draw(z1)--(z2);
\draw(z3)--(z2);
\draw(z2)--(z4);
\draw(z5)--(z6);
\draw(z5)--(z7);
\draw(z8)--(z9);
\draw(z9)--(z10);
\draw(z9)--(z11);
\draw(z12)--(z14);
\draw(z14)--(z13);
\draw(z15)--(z16);
\draw(z3)--(z17);
\draw(z5)--(z18);

\node[circle, fill=black, inner sep=0 pt, minimum size=0.1cm] (w2) at (8,8.2) [label=right:$ \pi(v)$]{};  
\node[circle, fill=black, inner sep=0 pt, minimum size=0.1cm] (w3) at (7.5,7.6) [label=left:$v$]{};
\node[circle, fill=black, inner sep=0 pt, minimum size=0.1cm] (w5) at (10,8.3) []{};  
\node[circle, fill=black, inner sep=0 pt, minimum size=0.1cm] (w8) at (7,11.2) []{};  
\node[circle, fill=black, inner sep=0 pt, minimum size=0.1cm] (w9) at (7,10.5) []{};  
\node[circle, fill=black, inner sep=0 pt, minimum size=0.1cm] (w10) at (6.5,9.9) []{};
\node[circle, fill=black, inner sep=0 pt, minimum size=0.1cm] (w11) at (7.5,9.9) []{}; 
\node[circle, fill=black, inner sep=0 pt, minimum size=0.1cm] (w12) at (8.5,9.9) []{};
\node[circle, fill=black, inner sep=0 pt, minimum size=0.1cm] (w13) at (9,9.9) []{};
\node[circle, fill=black, inner sep=0 pt, minimum size=0.1cm] (w14) at (8.75,10.6) []{};
\node[circle, fill=black, inner sep=0 pt, minimum size=0.1cm] (w15) at (10,10.6) []{};
\node[circle, fill=black, inner sep=0 pt, minimum size=0.1cm] (w16) at (10,10) []{};

\draw(w3)--(w2);
\draw(w8)--(w9);
\draw(w9)--(w10);
\draw(w9)--(w11);
\draw(w12)--(w14);
\draw(w14)--(w13);
\draw(w15)--(w16);

\path 
(w3) [orange] edge (w11)
(w2) [orange] edge (w11)
(z17) [orange] edge (z10)
(z17) [orange] edge (z12)
(y3) [orange] edge (y10)
(y2) [orange] edge (y13)
(x3) [orange] edge (x10)
(x4) [orange] edge (x12);

\node (c1) at (1.6,-1.5) [label=above:$\text{(c) Subsubcase} \  2.2.1$]{};
\node (c2) at (8.6,-1.5) [label=above:$\text{(d) Subsubcase} \  2.2.2$]{};

\node (c3) at (1.6,5.5) [label=above:$\text{(a) Case} \  1$]{};
\node (c4) at (8.6,5.5) [label=above:$\text{(b) Subcase} \  2.1$]{};

\node (d1) at (-1,1) [label=left:$H$]{};
\node (d2) at (-1,3.5) [label=left:$W_2$]{};

\node (d3) at (6,1) [label=left:$H$]{};
\node (d4) at (6,3.5) [label=left:$W_2$]{};

\node (d5) at (-1,8) [label=left:$H$]{};
\node (d6) at (-1,10.5) [label=left:$W_2$]{};

\node (d7) at (6,8) [label=left:$H$]{};
\node (d8) at (6,10.5) [label=left:$W_2$]{};

   \end{tikzpicture}
    \caption{Illustration of different cases in the proof of Theorem 1.}
    \label{fig:my_label}
\end{figure}

\noindent{\bf Case 1.}  Assume that  $v$ has at least two neighbours in $W_{2}$. 
Since Reduction EMMFVS 2 cannot be applied, we have that no two 
neighbours of $v$ belong to the same connected component of $G[W_{2}]$.
So, we can assume that all neighbours of $v$ belong to  different connected components of $G[W_{2}]$. See Figure \ref{fig:my_label}(a).
Now we branch by including $v$ in the solution in one branch and excluding it in the other branch. That is, we call the algorithm on instances $(G,W_{1}\cup \{v\},W_{2},k-1)$ and $(G,W_{1},W_{2}\cup \{v\},k)$. We check  minimality of the partial solution in every branch. If one of these branches returns a solution, then we conclude that $(G,W_{1},W_{2},k)$ is a yes-instance, otherwise $(G,W_{1},W_{2},k)$ is a no-instance.\\
 
\noindent{\bf Case 2.} Assume  that  $v$ has exactly 
one neighbour in $W_{2}$. Let $\pi(v)$ be the parent of $v$ in $H$.
We now have a number of subcases and subsubcases to consider.
Clearly, the degree of $\pi(v)$ cannot be one in $G-W_1$, otherwise Reduction EMMFVS 1 would have been applied. \\

\noindent{\bf Subcase 2.1.} 
Assume that the degree of $\pi(v)$ in $G-W_1$ is two. Then both $v$ and $\pi(v)$ are of degree two, and hence must have a common neighbour in $W_{2}$, 
otherwise Reduction EMMFVS 3 would have been applied. See Figure \ref{fig:my_label}(b).
Clearly, every solution of the {\sc Extension Max Min FVS} instance $(G,W_{1},W_{2},k)$ 
contains  either $v$ or $\pi(v)$. Note that, without loss of generality we can add $v$ inside the solution and keep $\pi(v)$ outside the solution. Therefore, we get reduced {\sc Extension Max Min FVS} instance $(G,W_{1}\cup \{v\},W_{2}\cup \{\pi(v)\},k-1)$. We check  minimality of this partial solution. \\

\noindent{\bf Subcase 2.2.} Assume that  the degree of $\pi(v)$ in $G-W_1$ is at least 3.
We split this subcase into two subsubcases.\\

\noindent{\bf Subsubcase 2.2.1.} 
Assume that $\pi(v)$ has no neighbours in $W_2$. As the degree of $\pi(v)$
in $G-W_1$ is at least three and it has no neighbours in $W_2$, it has at least two children. 
Without loss of generality, suppose $\pi(v)$ has two children $v$ and $v'$.
Observe that $v'$ is a leaf node, otherwise $v$ is not a deepest leaf in $H$. 
The degree of $v'$ cannot be one in $G-W_{1}$, otherwise Reduction EMMFVS \ref{Red1}
would have been applied. Also the degree of $v'$ cannot be more than one in $W_{2}$ as otherwise Case 1 will be applicable.
Therefore $v'$ has exactly one neighbour in $W_{2}$. 
Similarly, we can argue that $v$ has exactly one neighbour in $W_{2}$.
See Figure \ref{fig:my_label}(c).
In this case we make three branches by including one of $v$ and $v'$ in the solution and excluding $\{v,v',\pi(v)\}$ in the other branch.
That is, we get instances $(G, W_{1}\cup \{v\},W_{2},k-1),(G, W_{1} \cup \{v'\},W_{2},k-1)$ and $(G,W_{1},W_{2} \cup\{v,v',\pi(v)\},k)$. We check  minimality of the partial solution in every branch.
Notice that we have not considered a branch where $\pi(v)$ is inside the solution. 
For the sake of contradiction, assume that $\pi(v)$ is inside the solution $S$. 
We will prove that starting from $S$, we can construct another solution $S'$ such that $|S'|\geq |S|$ but $\pi(v) \not\in S'$. 
First we observe that if $\pi(v)$ is in $S$, 
none of its children are in $S$.
This is because each child of $\pi(v)$ has degree two in $G-W_1$ 
and one of its neighbours is $\pi(v)$.
That means every cycle that contains a child of $\pi(v)$ also contains 
$\pi(v)$. 
Next assuming that $\pi(v)$ is inside the solution $S$, there must exist a 
private cycle $C$ of $\pi(v)$.
Note that $C$ must contain at least one child of $\pi(v)$ as it does not 
have a neighbour in $W_{2}$. 
Without loss of generality, let that child  be $v$.
In this case we will replace $\pi(v)$ by $v$, that is, 
$S'=(S\setminus \{\pi(v)\})\cup \{v\} $. We observe that the solution will satisfy minimality condition because every cycle that is hit by $v$ is also hit by $\pi(v)$. \\  

\noindent{\bf Subsubcase 2.2.2.} Assume that $\pi(v)$ has at least one neighbour in $W_2$.
Note that $\pi(v)$ has at least one child. See Figure \ref{fig:my_label}(d).
In this case we get three branches by including one of $v$ and $\pi(v)$ in the solution and excluding $\{v,\pi(v)\}$ in the other branch. That is, we get instances $(G,W_{1} \cup \{v\},W_{2},k-1),(G, W_{1} \cup \{\pi(v)\},W_{2},k-1)$ and $(G,W_{1}, W_{2}\cup\{v,\pi(v)\},k)$. We check  minimality of the partial solution in every branch.\\


\begin{sloppypar} The algorithm stops when $k= 0$ or $H=\emptyset$. 
If $k=0$ at some leaf node in the search tree, then we conclude that 
 the given instance of  {\sc Extension Max Min FVS} is a 
yes-instance. Otherwise, it is a no-instance.
\emph{Note that at each branch of Case 1, Subcase 2.1, Subsubcase 2.2.1 and 2.2.2, we check the minimality of the partial solution, that is, we check if for every vertex $w\in W_{1}$ whether there exists a cycle containing $w$ in $G-\{W_{1}\setminus w\}$.} If the partial solution of a branch is not minimal, the we discard that branch. 
To estimate the running time of the algorithm for instance $I=(G,W_{1},W_{2},k)$, we use the measure
$\mu(I)$ as defined at the beginning of the proof. Observe that Reductions EMMFVS 1, EMMFVS 2, and EMMFVS 3 do not increase the measure. Now we see how $\mu(I)$ changes when we branch. 
In Case 1, when $v$ goes to the solution, $k$ decreases by $1$ and $\gamma(I)$ remains the same. Thus $\mu(I)$ decreases by 1. In the other branch, $v$ goes into $W_{2}$, then $k$ remains the same and $\gamma(I)$ decreases at least by 1. Thus $\mu(I)$ decreases at least by  1. Thus we have a branching vector $(1,\geq 1)$ for branching in Case 1. In Subcase 2.1, when $v$ or $\pi(v)$ goes to the solution, $k$ decreases by $1$ and $\gamma(I)$ remains the same. Thus $\mu(I)$ decreases by 1. In Subsubcase 2.2.1, clearly $\mu(I)$ decreases by 1 in the first and second branch as $k$ value decreases by  1. In the third branch, when we include $\{v, \pi(v), v'\}$ in $W_{2}$, $\gamma(I)$ drops at least by 1 and $k$ remains the same, therefore $\mu(I)$ decreases at least by 1. Thus, we have a branching vector $(1,1, \geq 1)$.
Similarly, we have a branching vector $(1,1,\geq 1)$ for Subsubcase 2.2.2.
As the maximum number of branches is $3$, the running time of our algorithm is $3^{\mu(I)}n^{\mathcal{O}(1)}$. Since we have $\mu(I)= k+\gamma(I)$, the running time of our algorithm is $3^{k+\gamma(I)}n^{\mathcal{O}(1)}$. \qed
\end{sloppypar} 
\subsection{An algorithm for {\sc Max Min FVS}}
 Given an input instance $(G,k)$, greedily find a minimal feedback vertex set $W$ of $G$.
If $|W|\geq k$, then $(G,k)$ is a yes-instance. Otherwise, we have 
a minimal feedback
vertex set $W$ of size at most $k-1$, that is, $\gamma(I)\leq k-1$ and the goal is to decide
whether $G$ has a minimal feedback vertex set $S$ of size at least $k$.
We do the following.
We guess the intersection of $S$ with $W$, that is, we guess the set $W_{1}=S\cap W$, and reduce parameter $k$ by $|W_{1}|$. For each guess of $W_{1}$,
we set $W_{2}=W\setminus W_{1}$ and solve {\sc Extension Max Min FVS} on the instance
$(G, W_{1},W_{2}, k-|W_{1}|)$. If for some guess, $G$ has  a minimal feedback vertex set $S$
 such that $W_{1} \subseteq S$, $W_2\cap S = \emptyset$ and 
$ |S\setminus W_{1}| \geq k-|W_1|$, then we conclude that the given instance of  {\sc Max Min FVS} is a 
yes-instance. Otherwise, 
we conclude that the given instance of  {\sc Max Min FVS} is a 
no-instance. The number of all guesses is bounded by $\sum\limits_{i=0}^{k-1} {k-1 \choose i}$.
We have an algorithm solving {\sc Extension  Max Min FVS} in time 
$3^{k+\gamma(I)}n^{\mathcal{O}(1)}=9^{k}n^{\mathcal{O}(1)}$ as $k+\gamma(I)\leq 2k-1$. Therefore we have an algorithm solving {\sc Max Min FVS}
in time 
$$ \sum\limits_{i=0}^{k-1} {k-1 \choose i} {9}^{k-i} n^{O(1)}=10^k n^{\mathcal{O}(1)}.$$
Thus we obtain Theorem~\ref{thmintro:mmfvs}. \\

 \section{FPT algorithm parameterized by vertex cover number}
In this section we prove that {\sc Max Min FVS} is  FPT when parameterized by vertex cover number $\vc(G)$.

\begin{proof}[The Proof of Theorem~\ref{thmintro:mmfvsvc}] 
If $G=(V,E)$ has a vertex $v$ of degree at most 1, remove it from the graph.
We find a vertex cover $C$ of size at most $\vc(G)$ of the reduced graph $G$. For our purpose, a standard branching algorithm with $O(2^{\vc(G)}n)$ running time is sufficient (see e.g. \cite{marekcygan}).
We denote by $I$ the independent set $V\setminus C$.
We next guess  $C_{in}=S\cap C$, where $S$ is a largest minimal FVS. 
There are at most $2^{\vc(G)}$ 
candidates for $C_{in}$ as each member of $C$ has two options: either in $C_{in}=S\cap C$ or $C_{out}=\bar{S} \cap C$.
Clearly, $ C_{out}=\bar{S} \cap C$ contains the vertices of $C$ which are outside the solution.  If $G[C_{out}]$ is not a forest then return that 
$C_{in}$ is a wrong guess and  reject it. 
So from now onwards we assume that  $G[C_{out}]$ is indeed a forest.
Next we check minimality of $C_{in}$, that is,  for each $v$ in $C_{in}$,   whether $ G[V\setminus C_{in} \cup \{v\}]$ has a cycle containing $v$. If the minimality of $C_{in}$ is
not satisfied then  return that 
$C_{in}$ is a wrong guess\\

\noindent{\it Outline of the algorithm:}  Given a guess $C_{in}$, our goal is to 
find a largest minimal FVS  containing $C_{in}$. We look for a set $Z\subseteq I$ of vertices which can be 
added to $C_{out}$ so that $G[C_{out}\cup Z]$ remains a forest and every vertex $v\in I\setminus Z$ has at least two neighbours in some component of $G[C_{out}\cup Z]$. 
This is why every vertex in $I\setminus Z$ must be included in the solution.
Finally the algorithm outputs $C_{in}\cup (I\setminus Z)$. \\

\noindent{\it Algorithm to find $Z$:} If $G-C_{in}$ has a vertex $v$ of degree at most 1, remove it from the graph.
If there is a vertex $v$ in $I$ such that $G[C_{out} \cup \{v\}]$ contains a cycle, then include $v$ in the solution. 
Suppose $S'$ is a minimal FVS such that $C_{in}\subseteq S'$.
We know $G-S'$ is a forest $\mathcal{F}$.
Suppose $\mathcal{F}$ has exactly  $q$ trees
 $T_1,T_2,\ldots, T_q$.  Note that the number of trees in $\mathcal{F}$ is at most $vc(G)$, that is, $q\leq vc(G)$. We guess a partition  $\mathcal{P}=\{P_{1},P_{2},\ldots,P_{q}\}$ of $C_{out}$. We say the partition $\mathcal{P}=\{P_{1},P_{2},\ldots,P_{q}\}$ corresponds to trees of $\mathcal{F}$, if 
 $P_i=C_{out} \cap V(T_i)$.
For each part $P_{i}$ there must exist a set $Z_{i} \subseteq I$ of vertices  such that 
$G[P_{i}\cup Z_{i}]=T_i$.
Otherwise  $\mathcal{P}$ is a wrong guess. Note that $Z=\bigcup_{i=1}^{q} {Z_i}$. There are at most $\vc(G)^{\vc(G)}$ candidates for $\mathcal{P}$  and we try out all  guesses. \\

\noindent{\it Algorithm to find $Z_i$:} Consider  $i$th part $P_i$  of 
$\mathcal{P}$. Note that $G[P_i]$ is  a collection trees.
Given  $P_i$, we want to have
a set $Z_i\subseteq I$ of vertices such that $G[P_{i}\cup Z_{i}]$ is a tree. 
This can happen in different ways. For example, it may be the case that only one 
vertex $z$ of $I$ connects all trees of $G[P_i]$ to form a single tree. It may be the case that we need $s_i>1$ vertices 
of $I$  to connect all trees of $G[P_i]$ to form a single tree. 
 We further guess a partition 
$\mathcal{P}_i=\{P_{i1},P_{i2},\ldots, P_{i{s_i}}\}$ of $P_i$ into $s_i$ parts.
For each $P_i$ there are at most ${\vc(G)}^{|P_i|}$ possible partitions.
Given a partition 
$\mathcal{P}_i=\{P_{i1},P_{i2},\ldots, P_{i{s_i}}\}$ of $P_i$ we want to have
a set
$Z_i=\{ z_{i1}, z_{i2}, \ldots, z_{is_i} \}$ of vertices such that
$z_{ij} \in I$ is adjacent to exactly one vertex of every tree in $P_{ij}$.  Thus $G[z_{ij}\cup P_{ij}]$ forms a 
 tree.  Next we guess how these  $s_i$ trees are joined to form a single  tree. 
We  need $s_i-1$  \emph{cross edges} of the form $(z_{ij}, v_{ik})$ where $v_{ik}\in P_{ik}$, $j\neq k$ to join $s_i$ trees. 
See Figure \ref{Z_i}.
\begin{figure}[ht]
    \centering
    \begin{tikzpicture}[scale=0.4]
    
\draw (0,0) ellipse (1.3cm and 0.4cm);

 \node(a) at (0,0.5) [label=above: $P_{i1}$]{};
  \node(b) at (3.5,0.5) [label=above: $P_{i2}$]{};
   \node(c) at (7,0.5) [label=above: $P_{i3}$]{};
    \node(d) at (10.5,0.5) [label=above: $P_{i4}$]{};

\node[circle,draw,fill=black, inner sep=0 pt, minimum size=0.1cm]	(p1) at (0,0) [label=left:]{};
\node[circle,draw,fill=black, inner sep=0 pt, minimum size=0.1cm]	(p2) at (-0.5,0) [label=left:]{};
v\node[circle,draw,fill=black, inner sep=0 pt, minimum size=0.1cm]	(p3) at (0.5,0) [label=left:]{};

\node[circle,draw,fill=black, inner sep=0 pt, minimum size=0.1cm]	(z1) at (0,-1.5) [label=below:$z_{i1}$]{};

\draw (3.5,0) ellipse (1.5cm and 0.4cm);

\node[circle,draw,fill=black, inner sep=0 pt, minimum size=0.1cm]	(p4) at (4.25,0) [label=left:]{};
\node[circle,draw,fill=black, inner sep=0 pt, minimum size=0.1cm]	(p5) at (3.75,0) [label=left:]{};
\node[circle,draw,fill=black, inner sep=0 pt, minimum size=0.1cm]	(p6) at (3.25,0) [label=left:]{};
v\node[circle,draw,fill=black, inner sep=0 pt, minimum size=0.1cm]	(p7) at (2.75,0) [label=left:]{};

\node[circle,draw,fill=black, inner sep=0 pt, minimum size=0.1cm]	(z2) at (3.5,-1.5) [label=below:$z_{i2}$]{};

\draw (7,0) ellipse (1.5cm and 0.4cm);

\node[circle,draw,fill=black, inner sep=0 pt, minimum size=0.1cm]	(p8) at (6.25,0) [label=left:]{};
\node[circle,draw,fill=black, inner sep=0 pt, minimum size=0.1cm]	(p9) at (7.75,0) [label=left:]{};
\node[circle,draw,fill=black, inner sep=0 pt, minimum size=0.1cm]	(p10) at (7.25,0) [label=left:]{};
v\node[circle,draw,fill=black, inner sep=0 pt, minimum size=0.1cm]	(p11) at (6.75,0) [label=left:]{};

\node[circle,draw,fill=black, inner sep=0 pt, minimum size=0.1cm]	(z3) at (7,-1.5) [label=below:$z_{i3}$]{};

\draw (10.5,0) ellipse (1.3cm and 0.4cm);
    
\node[circle,draw,fill=black, inner sep=0 pt, minimum size=0.1cm]	(p12) at (10.5,0) [label=left:]{};
\node[circle,draw,fill=black, inner sep=0 pt, minimum size=0.1cm]	(p13) at (11,0) [label=left:]{};
v\node[circle,draw,fill=black, inner sep=0 pt, minimum size=0.1cm]	(p14) at (10,0) [label=left:]{};    

\node[circle,draw,fill=black, inner sep=0 pt, minimum size=0.1cm]	(z4) at (10.5,-1.5) [label=below:$z_{i4}$]{};

\draw(z1)--(p1);
\draw(z1)--(p2);
\draw(z1)--(p3);

\draw(z2)--(p4);
\draw(z2)--(p5);
\draw(z2)--(p6);
\draw(z2)--(p7);

\draw(z3)--(p8);
\draw(z3)--(p9);
\draw(z3)--(p10);
\draw(z3)--(p11);

\draw(z4)--(p12);
\draw(z4)--(p13);
\draw(z4)--(p14);

\draw[red](z1)--(p5);
\draw[red](z1)--(p14);
\draw[red](z4)--(p10);

    \end{tikzpicture}
    \caption{An illustration for partition of $P_i$ into four parts $P_{i1}, P_{i2},P_{i3},P_{i4}$; cross edges $(z_{i1},v_{i2}),(z_{i1}, v_{i4}),(z_{i4},v_{i3})$ are shown in orange. For simplicity trees in  each part are singleton.}
    \label{Z_i}
\end{figure}
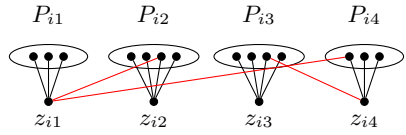
There are $s_i(s_i-1)$ cross edges of the form $(z_{ij}, v_{ik})$ where $v_{ik}\in P_{ik}$, $j\neq k$. Thus $s_i-1$ edges can be selected in at most $(s_i^2)^{s_i}$  many ways. 
So the total number of selections of cross edges for all $P_i$'s together is at most
$\prod_{i=1}^{q}{(s_i^2)^{s_i}} \leq \vc(G)^{2\vc(G)}$. Thus there are total $\vc(G)^{4\vc(G)}$ guesses for the structure of trees involving
vertices $V(P_i)\cup Z_i$ and cross edges for all $i$ combined.
A selection of $s_i-1$ cross edges   is  a valid selection if the chosen 
cross edges can join $s_i$ trees to form one tree.  For example, three cross edges (orange) in Figure \ref{Z_i} can join four trees to produce one tree; so these three orange edges form a valid
selection. Note that the vertices in $Z_i$ are abstract; now we look for 
candidates of $z_{ij}\in Z_i$ for all $j$, in the input graph. 
For a given guess, a vertex $x\in I$ is a  
\emph{candidate}  for $z_{ij}$ if and only if $x$ is only adjacent to exactly one vertex from each tree in $P_{ij}$ and also it is adjacent  to a vertex in $P_{ik}$ when $(z_{ij}, v_{ik})$ is a cross edge in the 
given guess. For a particular guess there  could be several candidates for $z_{ij}$.  We claim  that only one of the candidates of $z_{ij}$ can go to $Z_i$ and the rest go to the solution. 
Suppose there are two candidates $x_1$ and $x_2$ for $z_{ij}$. Without loss of generality 
assume that $x_1$ goes to $Z_i$, that is, $x_1$ does not go to the solution. 
Then we prove that $x_2$ must go to the solution. We consider two cases: \\

\noindent{\it Case 1:} Assume that $P_{ij}$ contains at least two trees. Then there is a cycle $C$ containing $x_{2}$ where the remaining  vertices of $C$ are
outside the solution; see Figure \ref{Z_i_1} (a).  Therefore $x_2$ must go to the solution in order to destroy $C$. \\

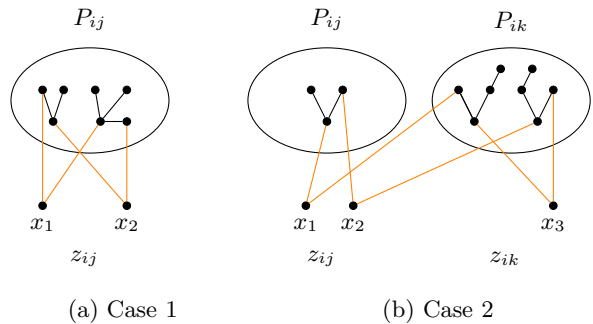
\begin{figure}[ht]
    \centering
    \begin{tikzpicture}[scale=0.7]
    
\draw (0,0) ellipse (1.5cm and 1.0cm);

 \node(a) at (0,1.0) [label=above:${P_{ij}}$]{};
  \node(b) at (4.5,1.0) [label=above:${P_{ij}}$]{};
   \node(c) at (8,1.0) [label=above:${P_{ik}}$]{};
   
 \node[circle,draw,fill=black, inner sep=0 pt, minimum size=0.1cm]	(p1) at (-0.7,-0.4) [label=left:]{};
\node[circle,draw,fill=black, inner sep=0 pt, minimum size=0.1cm]	(p4) at (-0.9,0.2) [label=left:]{};
\node[circle,draw,fill=black, inner sep=0 pt, minimum size=0.1cm]	(p5) at (-0.5,0.2) [label=left:]{};  
 \node[circle,draw,fill=black, inner sep=0 pt, minimum size=0.1cm]	(p2) at (0.2,-0.4) [label=left:]{};
 \node[circle,draw,fill=black, inner sep=0 pt, minimum size=0.1cm]	(p6) at (0.1,0.2) [label=left:]{};

\node[circle,draw,fill=black, inner sep=0 pt, minimum size=0.1cm]	(p3) at (0.7,-0.4) [label=left:]{};
\node[circle,draw,fill=black, inner sep=0 pt, minimum size=0.1cm]	(p7) at (0.7,0.2) [label=left:]{};

\node[circle,draw,fill=black, inner sep=0 pt, minimum size=0.1cm]	(x1) at (-0.9,-2) [label=below:$x_1$]{};
\node[circle,draw,fill=black, inner sep=0 pt, minimum size=0.1cm]	(x2) at (0.7,-2) [label=below:$x_2$]{};

   \draw(p1)--(p4);
   \draw(p1)--(p5);
   \draw(p2)--(p3);
   \draw(p2)--(p6);
   \draw(p2)--(p7);
   \draw[orange](x1)--(p2);
   \draw[orange](x1)--(p4);
   \draw[orange](x2)--(p1);
   \draw[orange](x2)--(p3);

    \draw (4.5,0) ellipse (1.5cm and 1.0cm); 
\node[circle,draw,fill=black, inner sep=0 pt, minimum size=0.1cm]	(r1) at (4.5,-0.4) [label=left:]{};
\node[circle,draw,fill=black, inner sep=0 pt, minimum size=0.1cm]	(r3) at (4.8,0.2) [label=left:]{};
\node[circle,draw,fill=black, inner sep=0 pt, minimum size=0.1cm]	(r2) at (4.2,0.2) [label=left:]{};

\node[circle,draw,fill=black, inner sep=0 pt, minimum size=0.1cm]	(y1) at (4.1,-2) [label=below:$x_1$]{};
\node[circle,draw,fill=black, inner sep=0 pt, minimum size=0.1cm]	(y2) at (5.0,-2) [label=below:$x_2$]{};
\node[circle,draw,fill=black, inner sep=0 pt, minimum size=0.1cm]	(y3) at (8.8,-2) [label=below:$x_3$]{};

\draw(r1)--(r2);
\draw(r1)--(r3);

    \draw (8.0,0) ellipse (1.5cm and 1.0cm);
    
\node[circle,draw,fill=black, inner sep=0 pt, minimum size=0.1cm]	(s1) at (7.3,-0.4) [label=left:]{};
\node[circle,draw,fill=black, inner sep=0 pt, minimum size=0.1cm]	(s3) at (7.0,0.2) [label=left:]{};
\node[circle,draw,fill=black, inner sep=0 pt, minimum size=0.1cm]	(s4) at (7.6,0.2) [label=left:]{};
\node[circle,draw,fill=black, inner sep=0 pt, minimum size=0.1cm]	(s7) at (7.8,0.6) [label=left:]{};

\node[circle,draw,fill=black, inner sep=0 pt, minimum size=0.1cm]	(s2) at (8.5,-0.4) [label=left:]{};
\node[circle,draw,fill=black, inner sep=0 pt, minimum size=0.1cm]	(s5) at (8.2,0.2) [label=left:]{};
\node[circle,draw,fill=black, inner sep=0 pt, minimum size=0.1cm]	(s6) at (8.8,0.2) [label=left:]{};
\node[circle,draw,fill=black, inner sep=0 pt, minimum size=0.1cm]	(s8) at (8.4,0.6) [label=left:]{};

\draw(s1)--(s3);
\draw(s1)--(s4);
\draw(s1)--(s3);
\draw(s4)--(s7);
\draw(s2)--(s5);
\draw(s2)--(s6);
\draw(s8)--(s5);    
\draw[orange](y1)--(r1);
\draw[orange](y1)--(s3);
\draw[orange](y2)--(r3);  
\draw[orange](y2)--(s2);  
\draw[orange](y3)--(s1);
\draw[orange](y3)--(s6);

\node(x0) at (0.5,-3) [label=left:$z_{ij}$]{};
\node(x0) at (5,-3) [label=left:$z_{ij}$]{};
\node(x0) at (8.5,-3) [label=left:$z_{ik}$]{};
\node(x0) at (2,-4) [label=left:$\text{(a) Case 1}$]{};
\node(x0) at (8,-4) [label=left:$\text{(b) Case 2}$]{};

   \end{tikzpicture}
    \caption{Illustration of Case 1 and Case 2.}
    \label{Z_i_1}
\end{figure}

\noindent{\it Case 2:} Assume that $P_{ij}$ contains exactly one tree. Recall that if $G-C_{in}$ has a vertex $v$ of degree at most 1,  then we remove it from the graph.
Therefore $x_1$ and $x_2$ have degree at least two.
Since $d(x_1), d(x_2)\geq 2$, both of them must have  a neighbour in some $P_{ik}$ where $j \neq k$, and $(z_{ij}, v_{ik})$ is a cross edge in the given guess. Therefore, the graph $G[V(P_{ij}) \cup V(P_{ik})\cup \{x_{1},x_{2}, x_3\}]$ has a cycle $C$ containing $x_{2}$ where
$x_3$ is a candidate vertex for $z_{ik}$ and $x_3$ is outside the solution.  
See Figure \ref{Z_i_1} (b). Note that $C$ contains $x_2$ where the remaining vertices are
outside the solution.
So $x_{2}$ must be inside the  solution in order to destroy $C$; this proves the claim.\\

\noindent So while choosing a candidate 
for $z_{ij}$, we only make sure that the remaining candidates which are going to the 
solution do not disturb the minimality of  $C_{in}$. If there is no such candidate then we
return that it is a wrong guess. 
Clearly, it takes  polynomial time to check if  $Z_{i}$ exists for all $i$. For a given guess
$C_{in}$, if  
 $Z=\bigcup_{i=1}^{q} {Z_i}$ exists in $I$ then we see that $C_{in} \cup (I \setminus Z) $ forms a minimal feedback vertex set containing $C_{in}$.

Given  $C_{in}$,  in order to compute 
$Z$ we consider at most ${\vc(G)}^{4\vc(G)}$ guesses. 
Thus given $C_{in}$ we can compute $Z$ in time 
${\vc(G)}^{4\vc(G)}n^{O(1)}$.  Given $C_{in}$
the above algorithm returns  either
a minimal FVS containing $C_{in}$ or returns $C_{in}$ is a wrong guess. Finally we consider the maximum size solution obtained over
all guesses. As there are $2^{\vc(G)}$ candidates for $C_{in}$, we can solve the 
problem in ${\vc(G)}^{5\vc(G)}n^{O(1)}$ time. 

\end{proof}

 \section{No polynomial kernel parameterized by $\vc(G)$}
 
 We proved that  {\sc Max Min FVS} is  FPT when parameterized by vertex cover number $\vc(G)$, and in this section we show kernelization hardness of
the problem.

 \begin{theorem}\label{pptGMDA}
 \mmfvs parameterized by $\vc(G)$ does not admit a polynomial 
compression unless coNP $\subseteq$ NP/poly. 
 \end{theorem}
 
\proof We give a polynomial parameter transformation (PPT) from the  {\sc Max Min VC} problem. Given an instance  $(G,k)$ of {\sc Max min VC}, we construct in polynomial time an instance $(G', k')$ of {\sc Max Min FVS} as follows. We start with the graph $G$. We add a new vertex $x$ and make it adjacent to every vertex of $G$. We add a  set  $V_{x}=\{x_{1},x_{2},\ldots,x_{n+3}\}$ of new vertices and make $x$ adjacent to 
every vertex of $V_x$.  Furthermore, we add a new vertex $y$ and make $y$ adjacent to every vertex of $V_{x}$. This completes the construction of $G'$. It is easy to see that $\vc(G') \leq \vc(G)+2$. Finally we set $k'=k+n+2$. We claim that $G$ contains a minimal vertex cover of size at least $k$ if and only if  $G'$ contains a minimal feedback vertex set of size at least $k'$.
\par Suppose $C$ is a minimal vertex cover in $G$ such that $|C|\geq k$. We observe that the set $C \cup (V_{x}\setminus \{x_{1}\})$  forms a minimal feedback vertex set  of size at least $k'$ in $G'$. Conversely, suppose that $G'$ has a minimal feedback vertex set $S$ of size at least $k'$. First we see that $x\not\in S$. This is true because if $x\in S$ then $(V_{x} \cup \{y\}) \cap S = \emptyset$ as the vertices  in $V_x\cup \{y\}$ are not part of any cycle in $G'- x$. This implies that $|S| \leq n+1$ which is a contradiction. Similarly, we can argue that $y \not\in S$. As we know that $\{x,y\} \cap S = \emptyset$, we must have $|S \cap V_{x}|=n+2$. This implies that $|S \cap V(G)|\geq k$. Now, we show that  $C = S \cap V(G)$ is a minimal vertex cover of $G$. It is easy to see that $C=S \cap V(G)$ is a vertex cover of $G$ otherwise $S$ will not be a feedback vertex set of $G'$. For the sake of contradiction assume that $C\setminus  \{x\}$ is also a vertex cover of $G$. In this case, we observe that $x$ is not part of any cycle in $G' - (S\setminus \{x\})$ which is a contradiction. Therefore $C$  is a minimal vertex cover of size at least $k$ in $G$. \qed


\section{An fpt-AS for {\sc Max Min FVS} parameterized by $\vc(G)$}

An \emph{fpt-approximation scheme (fpt-AS)} with parameterization 
$\kappa$ is an algorithm whose input is an instance $x\in I$ and an $\epsilon >0$, and
it produces a $(1-\epsilon)$-approximate solution in time 
$f(\epsilon ,\kappa(x))\cdot |x|^{O(1)}$ for some computable function $f$. 
In this section we prove Theorem~\ref{thmintro:mmfvsvcapx}.

\proof 
[The Proof of Theorem~\ref{thmintro:mmfvsvcapx}]
We present an $f(\epsilon, {\vc(G)})\cdot{n}^{O(1)}$ time 
algorithm that produces a $(1 - \epsilon)$-approximate solution for the problem, where 
$n$ is the number of vertices in the input graph $G$.
We assume that we have a minimum vertex cover $C$ of size $\vc(G)$  of the input graph $G=(V,E)$. 
We denote by $I$ the independent set $V\setminus C$.
Our goal here is to find a largest minimal FVS $S$ with $C_{in}=S\cap C$, where 
$C_{in} \subseteq C$ is given. That is,  we guess the intersection of $S$ with vertex cover $C$.  There are $2^{\vc(G)}$ 
possible guesses. Clearly, $ C_{out}=\bar{S} \cap C$ contains the vertices of $C$ which are outside the solution.  If $G[C_{out}]$ is not a forest then return that 
$C_{in}$ is a wrong guess and  reject it. 
So from now onwards we assume that  $G[C_{out}]$ is indeed a forest. We give some reduction rules to simplify the input instance. 
 

\begin{red}\label{red4}\rm 
If there is a vertex $u\in I$ with at most one neighbour in $C_{out}$, delete $u$.
\end{red}

\begin{red}\label{red5}\rm 
If there is a vertex $u\in I$  such that $G[C_{out}\cup \{u\}]$ contains a cycle, then include $u$ in the solution and  delete $u$. 
\end{red}

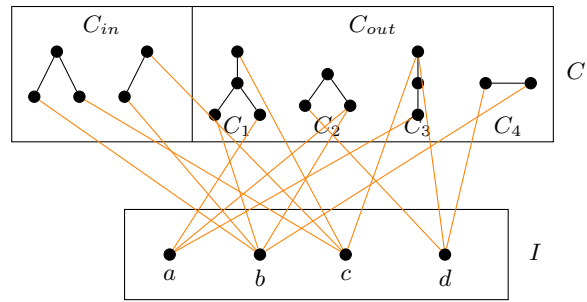
\begin{figure}[ht]
    \centering
   \begin{tikzpicture}[scale=0.6]
   \draw[black] (0,0) rectangle (12,3);
   \draw(4,0)--(4,3);
   \draw[black] (2.5,-3.5) rectangle (11,-1.5);
   \node[] at (2, 2) [label=above:$C_{in}$]{};
   \node[] at (8, 2) [label=above:$C_{out}$]{};
   \node[] at (12.5, 1) [label=above:$C$]{};
   \node[] at (11.6, -3) [label=above:$I$]{};
   \node[] at (5, -0.25) [label=above:$C_{1}$]{};
   \node[] at (7, -0.25) [label=above:$C_{2}$]{};
   \node[] at (9, -0.25) [label=above:$C_{3}$]{};
   \node[] at (11, -0.25) [label=above:$C_{4}$]{};
   
   \node[circle,draw,fill=black, inner sep=0 pt, minimum size=0.15cm]	(x1) at (1,2) [label]{};
   \node[circle,draw,fill=black, inner sep=0 pt, minimum size=0.15cm]	(x2) at (0.5,1) [label]{};
   \node[circle,draw,fill=black, inner sep=0 pt, minimum size=0.15cm]	(x3) at (1.5,1) [label]{};
   \node[circle,draw,fill=black, inner sep=0 pt, minimum size=0.15cm]	(x4) at (3,2) [label]{};
   \node[circle,draw,fill=black, inner sep=0 pt, minimum size=0.15cm]	(x5) at (2.5,1) [label]{};
   
   \node[circle,draw,fill=black, inner sep=0 pt, minimum size=0.15cm]	(y1) at (5,2) [label]{};
   \node[circle,draw,fill=black, inner sep=0 pt, minimum size=0.15cm]	(y2) at (5,1.3) [label]{};
   \node[circle,draw,fill=black, inner sep=0 pt, minimum size=0.15cm]	(y3) at (4.5,0.6) [label]{};
   \node[circle,draw,fill=black, inner sep=0 pt, minimum size=0.15cm]	(y4) at (5.5,0.6) [label]{};
   \node[circle,draw,fill=black, inner sep=0 pt, minimum size=0.15cm]	(y5) at (7,1.5) [label]{};
   \node[circle,draw,fill=black, inner sep=0 pt, minimum size=0.15cm]	(y6) at (6.5,0.8) [label]{};
   \node[circle,draw,fill=black, inner sep=0 pt, minimum size=0.15cm]	(y7) at (7.5,0.8) [label]{};
   \node[circle,draw,fill=black, inner sep=0 pt, minimum size=0.15cm]	(y8) at (9,2) [label]{};
   \node[circle,draw,fill=black, inner sep=0 pt, minimum size=0.15cm]	(y9) at (9,1.3) [label]{};
   \node[circle,draw,fill=black, inner sep=0 pt, minimum size=0.15cm]	(y10) at (9,0.6) [label]{};
   \node[circle,draw,fill=black, inner sep=0 pt, minimum size=0.15cm]	(y11) at (10.5,1.3) [label]{};
   \node[circle,draw,fill=black, inner sep=0 pt, minimum size=0.15cm]	(y12) at (11.5,1.3) [label]{};
   
   \draw(x1)--(x3);
   \draw(x1)--(x2);
   \draw(x4)--(x5);
   \draw(y1)--(y2);
   \draw(y2)--(y3);
   \draw(y2)--(y4);
\draw(y5)--(y6);
\draw(y5)--(y7);
\draw(y8)--(y9);
\draw(y10)--(y9);
\draw(y11)--(y12);

   \node[circle,draw,fill=black, inner sep=0 pt, minimum size=0.15cm]	(a) at (3.5,-2.5) [label=below:$a$]{};
   \node[circle,draw,fill=black, inner sep=0 pt, minimum size=0.15cm]	(b) at (5.5,-2.5) [label=below:$b$]{};
   \node[circle,draw,fill=black, inner sep=0 pt, minimum size=0.15cm]	(c) at (7.4,-2.5) [label=below:$c$]{};
   \node[circle,draw,fill=black, inner sep=0 pt, minimum size=0.15cm]	(d) at (9.6,-2.5) [label=below:$d$]{};
   
   \path
   (a) [orange] edge (y4)
   (a) [orange] edge (y7)
   (a) [orange] edge (y10)
   (b) [orange] edge (x2)
   (b) [orange] edge (x5)
   (b) [orange] edge (y3)
   (b) [orange] edge (y7)
   (b) [orange] edge (y12)
   (c) [orange] edge (x3)
   (c) [orange] edge (x4)
   (c) [orange] edge (y1)
   (c) [orange] edge (y8)
   (d) [orange] edge (y6)
   (d) [orange] edge (y8)
   (d) [orange] edge (y11);
   
   \end{tikzpicture}
   \caption{Here $Q(a)=\{C_1,C_2,C_3\}$, $ Q(b)=\{C_1,C_2,C_4\}$, $Q(c)=\{C_1,C_3\}$,   $Q(d)=\{C_2,C_3,C_4\}$. Note that 
   $S_a=\{b,c,d\}$.}
    \label{FPT-AS}
\end{figure}

The algorithm first applies Reductions EMMFVS \ref{Red1}, EMMFVS \ref{red4} and EMMFVS
\ref{red5} exhaustively. Every vertex $u\in I$ has at least two neighbours in 
$C_{out}$, otherwise Reduction EMMFVS \ref{red4} would have been applied. Since 
Reduction EMMFVS \ref{red5} cannot be applied, we have that no two neighbours of $u\in I$ belong to the same 
connected component of $G[C_{out}]$.
   On the reduced instance, we run the following greedy algorithm.
   Suppose that $C_{out}$ has connected components $C_1,C_2,..$. We say connected
   component $C_i$ is a neighbour of $u\in I $, that is $C_i\in Q(u)$, if 
   $G$ contains an edge $(u,v)$ for some $v\in C_i$.
We  pick an arbitrary vertex $u \in I$ and  define
$S_u=\Big\{x_i\in I~:~ |Q(u)\cap Q(x_i)|\geq 2   \Big\}$. See Figure \ref{FPT-AS}. 
Note that  if $u$ is not included in  $S$ 
then all the vertices of $S_u$ must be included in $S$. 
The intention is that if $|S_u|\geq 2$, then 
we prefer  not to include $u$ in $S$, and hence include all the vertices of  
$S_u$ in $S$ as it is a maximization problem. But while including the vertices 
of $S_u$ in the solution, 
we need to be careful about whether the inclusion of $S_u$ in the solution, disturbs 
the minimality property of  $C_{in}$.  This can be verified by
checking if each vertex in $C_{in}$ still has a private cycle after the 
inclusion of $S_u$ in the solution.
Based on the above observations, 
we propose the following algorithm. 
We  pick an arbitrary vertex $u \in I$, compute $S_u$ and check the minimality of
$C_{in}$ assuming $S_u$ is included in the solution.
If the minimality of $C_{in}$ is preserved then we set $S=S\cup S_u$, $I=I\setminus (S_u \cup \{u\})$, 
and $C_{out}=C_{out} \cup \{u\}$, that is,
we include $S_u$ in the solution and move $u$ to $C_{out}$. 
As $u$ has neighbours in at least two connected components of 
$G[C_{out}]$,  when we move $u$ to $C_{out}$, the number of components 
in $G[C_{out}]$ drops by at least 1. The algorithm again applies Reductions  EMMFVS \ref{red4} and EMMFVS
\ref{red5} exhaustively as $C_{out}$ has been modified.
On the other hand, if the minimality of $C_{in}$ is not preserved then we set $S=S\cup \{u\}$ and 
$I=I\setminus \{u\}$.
We repeat the above  until 
$I$ becomes empty.

There are $2^{\vc(G)} $ candidates for $C_{in}$; for each guess 
the above algorithm returns  
a minimal FVS  and finally we consider the maximum size solution obtained over
all guesses.  Suppose the algorithm outputs $S$. Let $S_{opt}$ be an
optimum solution.
We claim that $|S|\geq |S_{opt}| - {\vc(G)}$. 
Let $C_{in}=C\cap S_{opt}$.
Clearly, we have $|S_{opt}| \leq |C_{in}| + |I|$. 
Recall that the greedy algorithm adds all vertices from $I$ to the 
solution except the ones that are moved to $C_{out}$. 
We claim that there are at most ${\vc(G)}$ vertices from $I$ that are 
moved to $C_{out}$ and therefore not added to the solution. 
Every time a vertex is moved to $C_{out}$, the number of connected components in  $G[C_{out}]$ drops by at least one. After moving ${\vc(G)}-1$ vertices,  
the number of connected components  in $G[C_{out}]$ becomes one. 
Therefore, we have moved at most ${\vc(G)}$ vertices to $C_{out}$. This proves that $|S|\geq |C_{in}|+ I-{\vc(G)} \geq |S_{opt}| - {\vc(G)}$.

\par Next, we propose an FPT approximation scheme. 
Given an input graph $G$, we first ask whether $G$ has a minimal FVS of 
size at least $\frac{\vc(G)}{\epsilon}$. Note that this can be 
answered in time $10^{\frac{{\vc(G)}}{\epsilon}} n^{O(1)}$ using the FPT algorithm proposed in Section \ref{FPT-Algo}. If this is a no-instance 
then we can find ${\sf opt}(G)$ in the same time by repetitively using the FPT algorithm. If this is a yes-instance, then  obviously   $ {\sf opt}(G)\geq \frac{{\vc(G)}}{\epsilon}$.
Hence the value of the constructed solution $S$ is at least
\begin{equation*}
    \frac{{\sf opt}(G) - {\vc(G)}}{{\sf opt}(G)} \geq 1 - \frac{{\vc(G)}}{{\sf opt}(G)} \geq 1-\epsilon
\end{equation*}
times the optimum. That is, 
the constructed solution $S$ is a $(1-\epsilon)$-approximate solution. \qed

\section{Conclusion and Open Problems}
In this paper, we have studied \mmfvs parameterized by the solution size and the vertex cover number of the input graph. We gave a single exponential time algorithm for \mmfvs parameterized by solution size and a $\csO(2^{\cO(\vc(G) \log \vc(G))})$ time algorithm parameterized by $\vc(G)$. Finally we proposed  an FPT-AS
parameterized by $\vc(G)$ with better running time $2^{\mathcal{O}\big(\frac{\vc(G)}{\epsilon}\big)} n^{\mathcal{O}(1)}$.
We list some nice problems that emerge from the results here: can our algorithm parameterized 
by $\vc(G)$ be made $c^{\cO(\vc(G))}$ time algorithm, for a fixed constant $c$?
A $2^{\mathcal{O}(\omega^{\cO(1)})}n^{\mathcal{O}(1)}$ time algorithm seems possible where $\omega$ is the treewidth of the input graph. Therefore, it would be interesting to see if the idea in Theorem \ref{thmintro:mmfvsvc} can be extended to study \mmfvs parameterized by 
${\sf fvs}(G)$.

\bibliographystyle{abbrv}
\bibliography{bibliography}

\end{document}